\documentclass[10pt]{article}
\usepackage{wws2025} 
\usepackage{url}
\usepackage{latexsym}
\usepackage{amsmath, amsthm, amsfonts}
\usepackage{algorithm, algorithmic}  
\usepackage{graphicx}
\usepackage{lipsum} 
\usepackage{fancyhdr}
\usepackage[headheight=1in,margin=1in]{geometry}
\usepackage{hyperref}
\usepackage[T1]{fontenc}

\title{Utilizing citation index and synthetic quality measure to compare Wikipedia languages across various topics}

\author{
  Włodzimierz Lewoniewski \\
  Poznań University \\of Economics and Business \\\And
 Krzysztof Węcel \\
  Poznań University \\of Economics and Business \\\And
  Witold Abramowicz \\
  Poznań University \\ of Economics and Business \\}

\begin{document}
\maketitle
\thispagestyle{fancy}

\begin{abstract}
This study presents a comparative analysis of 55 Wikipedia language editions employing a citation index alongside a synthetic quality measure. Specifically, we identified the most significant Wikipedia articles within distinct topical areas, selecting the top 10, top 25, and top 100 most cited articles in each topic and language version. This index was built on the basis of wikilinks between Wikipedia articles in each language version and in order to do that we processed 6.6 billion page-to-page link records. Next, we used a quality score for each Wikipedia article - a synthetic measure scaled from 0 to 100. This approach enabled quality comparison of Wikipedia articles even between language versions with different quality grading schemes. Our results highlight disparities among Wikipedia language editions, revealing strengths and gaps in content coverage and quality across topics.
\end{abstract}

\section*{Introduction}
Despite Wikipedia’s global reach, large disparities persist in both coverage and quality across its over 300 language editions. Quantifying these gaps can be essential for targeted community efforts. In this paper, we employ citation-based rankings and synthetic quality measures to assess and compare the quality of Wikipedia articles across 55 language editions and 18 selected topics. By focusing on the most cited articles within each topic, we provide insights into the relative strengths and weaknesses of various Wikipedia languages. Citation counts can capture collective editorial attention, but alone they do not guarantee depth or verifiability. By pairing a link-based prominence index with a normalized, multi-feature quality score, we gain a more nuanced view of which articles are both widely referenced and substantively mature.

In order to analyze the data, we prepared algorithms that extract different features from the Wikimedia dumps as of 1 August 2024.

\section*{Citation index}
In order to get the citation index for each of the 47 million articles in 55 language versions of Wikipedia, we used dumps with wiki page-to-page link records (for example, for English Wikipedia it was ''enwiki-20240801-pagelinks.sql.gz''). In total, we extracted and analyzed 6.6 billion wiki page-to-page link records. To construct our citation index, we considered only incoming links from the Wikipedia articles - 4.1 billion links from the main namespace (ns 0). Additionally, we published the dataset containing the calculated citation indexes \footnote{\url{https://huggingface.co/datasets/lewoniewski/wikipedia-citation-index}}.

\section*{Quality measure}
In prior research \cite{Wecel2015,Lewoniewski2017i,lewoniewski2020}, we developed a synthetic measure designed to integrate multiple key characteristics of Wikipedia articles, enabling an evaluation of their quality on a unified scale ranging from 0 to 100. The synthetic measure incorporates five essential features: article length, the total number of references, reference density, the count of images, and the number of sections within an article. Given that each Wikipedia language edition maintains distinct standards for designating top-quality articles -- analogous to the ''Featured Articles'' used by English Wikipedia -- we normalize these features relative to each language’s benchmarks, based on median values of the best-rated articles within that language edition. Normalization is executed by comparing each article's feature value against the median threshold from the respective language. Values meeting or exceeding the median are assigned 100 points, while values below the median are proportionally scaled. Therefore, we first counted the normalized metrics average (NMA) by the following formula:
\begin{equation}
NMA=\frac{1}{c}\sum_{i=1}^c \hat{m}_{i},
\end{equation}
where~$\hat{m}_{i}$ is a normalized measure~$m_{i}$ and~$c$ is the number of measures.

Next we took into account the number of quality-flaw templates \footnote{For example, some of the quality-flaw templates of English Wikipedia are described at \url{https://en.wikipedia.org/wiki/Wikipedia:Template_index/Disputes}} ({QFT}) in the considered article (if~they~existed). Those templates are standardized banners and inline tags used to inform readers about specific problems in articles, such as  lack of sources, neutrality issues, or factual accuracy. After including QFT, our final formula for the quality measure reads as follows:
\begin{equation}
Quality Score = NMA \cdot (1 - 0.05\cdot QFT).
\end{equation}

The dataset with the quality score for each Wikipedia article was also published online \footnote{\url{https://www.kaggle.com/datasets/lewoniewski/quality-of-wikipedia-articles-by-wikirank}}. The implemented version of the quality score is available on WikiRank \footnote{\url{https://wikirank.net}}.

\section*{Topic identification}
Based on our previous approach \cite{lewoniewski2020} we align articles on various topics using their connections to appropriate Wikidata items. For example, to align an article to the "city" category we select those related Wikidata items that have in ''instance-of'' (P31) property such values as ''city'' (Q515), ''big city'' (Q1549591), ''city with powiat rights'' (Q925381) and others. Due to space limitations, we confined our analysis to 18 distinct topics (see the figure \ref{figure}).

\section*{Results and Discussion}

\begin{figure*}[ht]
\begin{center}
\centerline{\includegraphics[width=\textwidth]{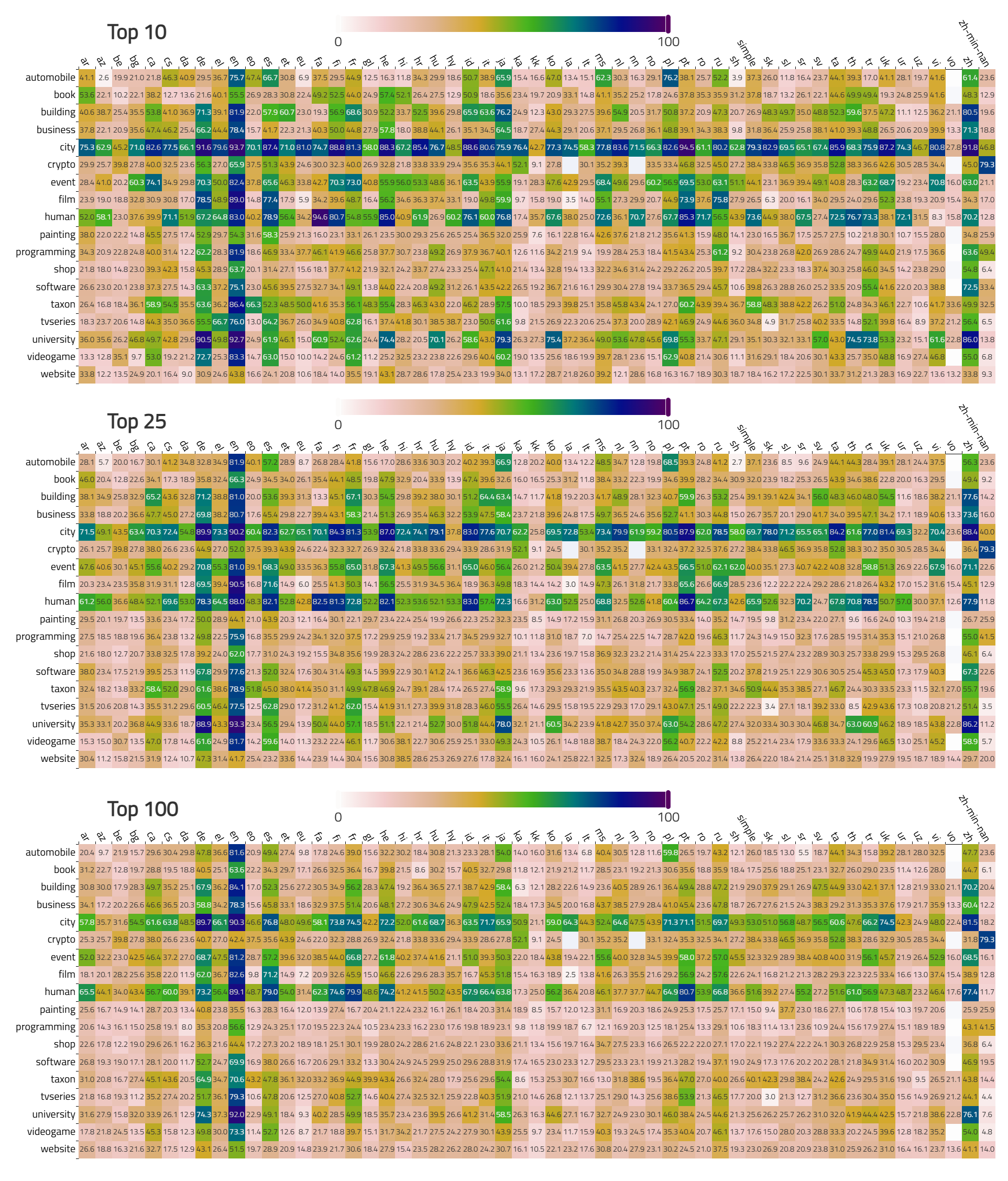}}
\caption{Average quality score across Wikipedia languages and topics within the Top 10, Top 25 and Top 100 most cited articles. Interactive version of the charts is available at: \url{https://data.lewoniewski.info/wikiworkshop2025}}
\label{figure}
\end{center}
\end{figure*}

First, let’s consider the results from the Top 10 analysis presented in the figure \ref{figure}. English Wikipedia (en) and German Wikipedia (de) consistently exhibit the highest quality across nearly all topics, indicating a strong alignment between citation prominence and article comprehensiveness. English Wikipedia particularly excels in articles about cities (93.73), films (89.01), taxon (86.37), and human-related content (82.98), emphasizing its extensive depth and quality in mainstream encyclopedic topics. Languages such as Catalan (ca), German (de), Spanish (es), Korean (ko), and Chinese (zh) also demonstrate consistently higher scores across various categories, especially in topics like "city", "university", and "events", indicating robust coverage and comprehensive content development in these Wikipedias. The analysis suggests that universally, the "city" category shows the highest average quality across most languages, followed by "human" and "university", whereas topics such as "videogame", "painting", and "automobile" typically score lower, indicating more limited international coverage or less prioritization.

Results from the Top 25 most cited articles showed that German (de), Spanish (es), and Chinese (zh) consistently maintain high-quality scores across multiple topics. English Wikipedia continues to lead, especially in categories like "city", "film", "human", and "university", maintaining scores typically above 80. In general, when comparing the Top 25 to the Top 10, average quality scores decreased slightly across several languages and topics. 

Finally, let's analyze results from the Top 100 most cited articles. Overall, the highest-quality scores are still predominantly observed in widely developed language editions such as English (en), German (de), Spanish (es), and Chinese (zh), though quality tends to decrease when a larger set of articles (Top 100) is evaluated. English Wikipedia maintains strong leadership in numerous topics, particularly in "city", "human", and "university", although some categories, like "programming" and "painting", exhibit a notable drop in quality compared to the Top 25. For less developed language versions, including Azerbaijani (az), Belarusian (be), and Lithuanian (lt), the expansion to the Top 100 consistently results in lower average quality scores across most topics, highlighting limited content depth beyond the most cited articles. Azerbaijani Wikipedia, for instance, shows substantial declines, particularly in the "city" category, dropping from 62.91 (Top 10) to 35.7 (Top 100). Arabic Wikipedia (ar) demonstrates significant fluctuations, with categories like "event" increasing progressively from 28.42 to 51.97. 

In general, the quality disparities observed between the Top 10, Top 25, and Top 100 analyses suggest the presence of concentrated efforts on highly cited articles, while moderately and less cited articles often lack comparable editorial attention, particularly in smaller languages.

This study represents a foundational step toward a more comprehensive analysis of multilingual Wikipedia diversity. In future works, we plan to expand our scope to encompass a broader range of topical categories and an even larger set of language editions. We will also integrate additional metrics, such as the page views statistics to capture the most popular articles, and the number of unique editors to gauge author diversity. 

\bibliographystyle{wws2025} 
\bibliography{references}

\begin{thebibliography}{}

\bibitem[\protect\citename{Lewoniewski \bgroup et al.\egroup
  }2017]{Lewoniewski2017i}
Włodzimierz Lewoniewski, Krzysztof Węcel, and Witold Abramowicz.
\newblock 2017.
\newblock {Relative Quality and Popularity Evaluation of Multilingual Wikipedia
  Articles}.
\newblock {\em Informatics}, 4(4).

\bibitem[\protect\citename{Lewoniewski \bgroup et al.\egroup
  }2019]{lewoniewski2020}
Włodzimierz Lewoniewski, Krzysztof Węcel, and Witold Abramowicz.
\newblock 2019.
\newblock {Multilingual Ranking of Wikipedia Articles with Quality and
  Popularity Assessment in Different Topics}.
\newblock {\em Computers}, 8(3).

\bibitem[\protect\citename{Węcel and Lewoniewski}2015]{Wecel2015}
Krzysztof Węcel and Włodzimierz Lewoniewski.
\newblock 2015.
\newblock {Modelling the Quality of Attributes in Wikipedia Infoboxes}.
\newblock In Witold Abramowicz, editor, {\em {B}usiness {I}nformation {S}ystems
  {W}orkshops}, volume 228 of {\em Lecture Notes in Business Information
  Processing}, pages 308--320. Springer International Publishing.

\end{thebibliography}

\clearpage
\newpage

\end{document}